\documentclass[aps,pre,showpacs,groupedaddress]{revtex4}
\usepackage{graphicx}
\begin{document}

\title{Growing Correlation Length on Cooling Below the Onset of Caging in a 
Simulated Glass-Forming Liquid}   
    
\author{N.~La\v{c}evi\'{c}$^{1,2,3}$, F.W.~Starr$^2$,
 T.B.~Schr\o der$^{2,4}$,
 V.N.~Novikov$^{2,}$\footnote{Permanent address: Institute of
 Automation and Electrometry, Russian Academy of Science, Novosibirsk,
 630090, Russia} and S.C.~Glotzer$^{1,2}$ }

\affiliation{${}^1$ Departments of Chemical Engineering and 
Materials Science and Engineering, \\ University of Michigan, Ann
Arbor, MI 48109, USA}

\affiliation{${}^2$ Center for Theoretical and Computational Materials
Science and Polymers Division, \\ National Institute of Standards and
Technology, Gaithersburg, MD 20899, USA}

\affiliation{${}^3$ Department of Physics and Astronomy, 
Johns Hopkins University \\ Baltimore, MD 21210, USA}

\affiliation{${}^4$ IMFUFA, Roskilde University, DK-4000, Denmark}

\date{June $18$, $2002$ }

\begin{abstract} 
We present a calculation of a fourth-order, time-dependent density
correlation function that measures higher-order spatiotemporall
correlations of the density of a liquid. From molecular dynamics
simulations of a glass-forming Lennard-Jones liquid, we find that the
characteristic length scale of this function has a maximum as a
function of time which increases steadily beyond the characteristic
length of the static pair correlation function $g(r)$ in the temperature
range approaching the mode coupling temperature from above.
\end{abstract}

\maketitle

Relaxation in liquids near their glass transition involves the
correlated motion of groups of neighboring 
particles~\cite{sillreview,edigerreview,scg}. 
This correlated motion results in spatially heterogeneous dynamics, 
which becomes increasingly heterogeneous as the liquid is cooled. 
Much remains to be understood regarding the nature of this heterogeneity, 
and how to best measure and quantify it. 
The traditional two-point, time-dependent, van Hove density
correlation function $G(r,t)$, provides information about the 
transient ``caging'' of particles on cooling \cite{gotze}, but does
not provide local information about correlated motion and
dynamical heterogeneity. In particular, the static correlation length
associated with two-point density fluctuations remains relatively
constant upon cooling \cite{grest,lmnpst_wiltzius}.  Instead, other correlation
functions, which involve, e.g., spatial correlations of the
displacements of particles in the liquid, and other measures of 
correlated motion, have been used to demonstrate the heterogeneous 
nature of the liquid dynamics in molecular dynamics (MD) simulations
\cite{kdppg-gssg,domh-sim-l}. These
``measures'' are readily accessible in colloidal suspensions, where
microscopy provides detailed information on particle trajectories
similar to the information obtained from MD simulations.  Indeed, such
experiments have confirmed simulation predictions of
increasingly heterogeneous dynamics near the glass transition
\cite{kegel-weeks}. 

In this paper, we evaluate a fourth-order, time-dependent density
correlation function $g_4(r,t)$ that is more sensitive to spatially
heterogeneous dynamics than $G(r,t)$.  This four-point function was
first investigated in supercooled liquids by Dasgupta et
al. \cite{dasgupta91}, but they did not detect a growing correlation
length in their simulations. Recently, it was shown \cite{dfpg,gns}
that it is possible to define a generalized, time-dependent
susceptibility $\chi_4(t)$ which (i) is proportional to the volume
integral of $g_4(r,t)$ in the same way that the isothermal
compressibility is related to the volume integral of the static pair
correlation function \cite{hansen-mcdonald}, (ii) is non-zero in the
caging regime, and (iii) increases with decreasing $T$.  While this
indirectly suggests the presence of a growing correlation length,
neither $g_4(r,t)$ nor its range $\xi_4(t)$ was calculated in those
works.  Here we calculate $g_4(r,t)$ in a simulation of $8000$
Lennard-Jones (LJ) particles, and show that $\xi_4(t)$ grows slowly
but steadily beyond the correlation length $\xi$ of the static pair
correlation function $g(r)$ as temperature $T$ is decreased from the
onset of caging towards the mode coupling temperature $T_{{\rm MCT}}$
\cite{gotze}.

We first briefly review the general theoretical framework, some of
which was previously discussed in Refs.~\cite{dfpg,gns,Overlaps},
and extend it to obtain a form for $g_4(r,t)$ suitable for calculation
in our simulations.  Consider a liquid of $N$ particles occupying a
volume $V$, with density $\rho(\textbf{r},t) = \sum_{i=1}^N \delta
(\textbf{r}-\textbf{r}_i(t))$.
Extending an idea originally proposed for spin glasses \cite{parisi-spin-g}, 
one may construct a time-dependent `order parameter' that compares the 
liquid configuration at two different times 
\cite{dfpg,gns,Overlaps}:
\begin{eqnarray}
\label{Qtot}    
Q(t) = \int d\textbf{r}_1 d\textbf{r}_2 \rho(\textbf{r}_1,0)
\rho(\textbf{r}_2,t) w(|\textbf{r}_1 - \textbf{r}_2|) 
 = \sum_{i=1}^{N} \sum_{j=1}^{N} w(|\textbf{r}_{i}(0) - \textbf{r}_j(t)|).
\end{eqnarray}
Here, $\textbf{r}_i$ in the second equality refers to the position
of particle $i$, and $w(|\textbf{r}_i - \textbf{r}_j|)$
is an ``overlap'' function which is unity for $|\textbf{r}_i -\textbf{r}_j| 
\leq a$ and zero otherwise, where the parameter $a$ is associated with the 
typical vibrational amplitude of the particles\cite{dfpg,gns}. 
For the present work, we take $a=0.3$ particle diameters 
as in Refs.~\cite{dfpg,gns}, since this value maximizes the effect studied
here~\cite{lssng}. As defined, $Q(t)$ is the number 
of ``overlapping'' particles when configurations of the system 
at $t=0$ and at a later time $t$ 
are compared; that is, $Q(t)$ counts the number of particles that either
remain within a distance $a$ of their original position, or are replaced
(within a distance $a$) by another particle in an interval $t$.

The fluctuations in $Q(t)$ are described by a generalized 
susceptibility
\begin{eqnarray}
\chi_4(t) = \frac{\beta V}{N^2}
\Big( \langle Q^2(t) \rangle - \langle Q(t) \rangle^2 \Big),
\label{chi4}
\end{eqnarray}
where $\beta = 1/k_BT$, $k_B$ is Boltzmann's constant, 
and $\langle \cdots \rangle$
indicates an ensemble average as in Ref.~\cite{gns}. 
Note that at very early times, when 
$Q(t) = N$ because no particle has yet moved beyond a 
distance $a$, $\chi_4(t)$ 
is identical to the isothermal compressibility $\kappa_T$ \cite{hes}. 
Substituting Eq.~(\ref{Qtot}) into Eq.~(\ref{chi4}), we obtain
\begin{eqnarray}
\label{chi4G}
\chi_4(t) = \frac{\beta V}{N^2} \int d\textbf{r}_1 d\textbf{r}_2
d\textbf{r}_3 d\textbf{r}_4 \> G_4(\textbf{r}_1, \textbf{r}_2, \textbf{r}_3,
\textbf{r}_4, t),
\end{eqnarray}
where
\begin{eqnarray}
\label{Grs}
G_4(\textbf{r}_1, \textbf{r}_2, \textbf{r}_3,\textbf{r}_4, t) &=&
\langle \rho(\textbf{r}_1,0)\rho(\textbf{r}_2,t)w(|\textbf{r}_1 - \textbf{r}_2|)\rho(\textbf{r}_3,0)
\rho(\textbf{r}_4,t) w(|\textbf{r}_3 - \textbf{r}_4|) \rangle \nonumber \\ &-& \langle
\rho(\textbf{r}_1,0)\rho(\textbf{r}_2,t)w(|\textbf{r}_1 - \textbf{r}_2|) \rangle \langle
\rho(\textbf{r}_3,0) \rho(\textbf{r}_4,t)w(|\textbf{r}_3 - \textbf{r}_4|) \rangle.
\end{eqnarray}

As written, $G_4$ is a function of four spatial and two temporal coordinates, 
constrained in a particular way by the overlap functions.
To investigate the behavior of $G_4$, we consider a function $g_4(r,t)$ such 
that $\chi_4(t) = \beta \int d{\bf r} \ g_4(r,t)$. We integrate first 
over $\textbf{r}_2$ and $\textbf{r}_4$ in Eq.~(\ref{chi4G}), define 
$\textbf{r} \equiv \textbf{r}_3 - \textbf{r}_1$, and then integrate over 
$\textbf{r}_3$ to obtain
\begin{eqnarray}
\label{eqn15}
   g_4(r,t) = \Big \langle \frac{1}{N \rho} \sum_{ijkl} 
    \delta(\textbf{r} - \textbf{r}_k(0) + \textbf{r}_i(0)) w(|\textbf{r}_i(0) - \textbf{r}_j(t)|) 
    w(|\textbf{r}_k(0) - \textbf{r}_l(t)|) \Big \rangle -
     \Big \langle \frac{Q(t)}{N} \Big \rangle^{2}
\end{eqnarray}
We investigate the behavior of $g_4(r,t)$, which is the angular averaged 
function of a single variable $\textbf{r}$.
With the above choice of integration variables, $g_4(r,t)$ describes spatial 
correlations between overlapping particles at the initial time 
(using information at time $t$ to label the overlapping particles). 
The first term in $g_4(r,t)$ is a pair correlation function restricted
to the subset of overlapping particles, $g^{\rm{ol}}_4(r,t)$. 
The second term represents 
the ``bulk'' probability of any two particles overlapping. We can rewrite $g_4(r,t)$ 
as
\begin{eqnarray}
\label{eqn15a}
  g_4(r,t) &=& g_4^{\rm{ol}}(r,t) -  \Big \langle \frac{Q(t)}{N}  \Big \rangle^2  = 
                \Big \langle \frac{Q(t)}{N}  \Big \rangle^2 
                \Big [\frac{g_4^{\rm{ol}}(r,t)}{ \langle \frac{Q(t)}{N} \rangle^2} -1 \Big]
          \nonumber \\
         &\equiv&  \Big \langle \frac{Q(t)}{N} \Big \rangle^2 g^{*}_4(r,t).
\end{eqnarray}
Since $ \langle \frac{Q(t)}{N} \rangle $ is a function of time,
information about spatial correlations is contained in $g^{*}_4(r,t)
$, which we investigate in the rest of the paper. In comparing $g_4(r,t)$ with other
correlation functions used to study glass-forming liquids, we note that neutron 
scattering studies, such as those by Colmenero and Richter and coworkers on polymer
systems \cite{arbe}, measure at most two-point spatiotemporal density correlation functions.
$4$-D NMR methods used to probe dynamical heterogeneity measure multiple
time correlation functions \cite{spiess-schmidt-rohr}. In contrast to those,
$g_4(r,t)$ contains additional, higher order spatiotemporal information.

To evaluate $g^{*}_4(r,t)$, we perform MD simulations 
of a model LJ glass-forming liquid. The system is a three-dimensional
(50:50) binary mixture of $8000$ particles interacting via LJ
interaction parameters~\cite{gns,units}. We simulate eight state
points in the microcanonical ensemble at fixed density $\rho=1.29$, 
in a temperature range $0.59 - 2.0$; we estimate 
$T_{{\rm MCT}} = 0.57 \pm 0.02$ and the Vogel-Fulcher temperature 
$T_0=0.48 \pm 0.02$. The error bars are confidence intervals obtained
as a result of fitting $\tau_{\rm{\alpha}}(T)$ to a power law and 
exponential form, respectively.
The onset of caging and non-exponential relaxation of
the intermediate scattering function occur at $T_{\rm cage} \approx 1.0$. 
Our simulations thus span a range that includes  the onset of caging and
the initial slowing down of the liquid approaching $T_{\rm MCT}$. 
The isochore along which the simulations are performed was chosen 
to reproduce simulations of a smaller system size performed earlier, 
which showed (i) that this model exhibits spatially heterogeneous 
dynamics\cite{gns}, and (ii) that transitions between inherent 
structures close to $T_{\rm MCT}$ occur through the collective, 
quasi-one-dimensional motion of strings of particles \cite{ssdg}. 
All data evaluated in the present work are in thermodynamic 
equilibrium above the glass transition temperature. 

\begin{figure}[tbp] 
\begin{center}
\includegraphics[clip,width=15cm]{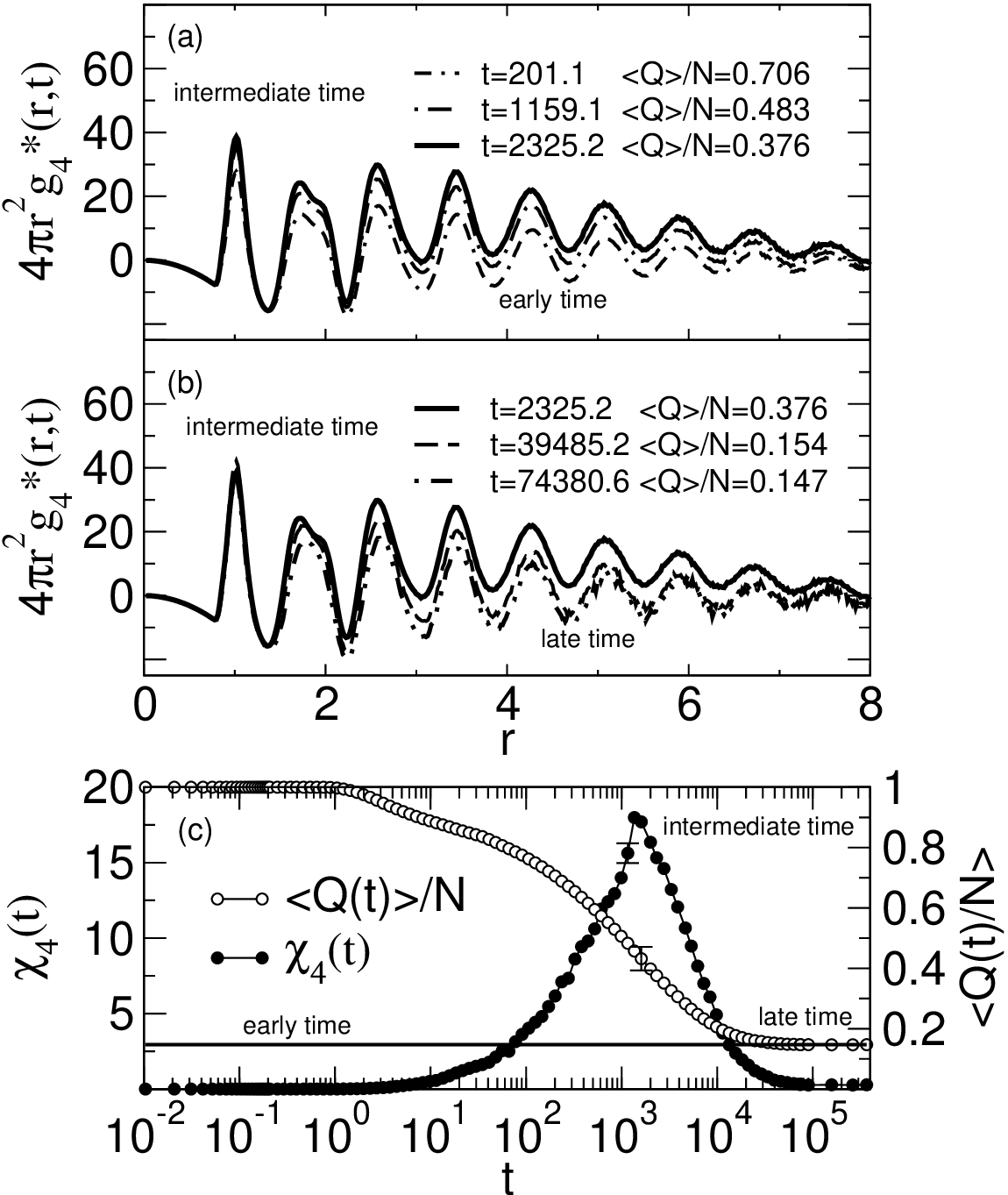}
\end{center}
\caption{Time dependence of $g^{*}_4(r,t)$ at $T=0.60$,
showing (a) the amplitude and range of $g^{*}_4(r,t)$
growing in time and (b) the amplitude and range of 
$g^{*}_4(r,t)$ decaying in time.
The fraction $\langle \frac{Q(t)}{N} \rangle $ indicates the average fraction
of overlapping particles present at time $t$. (c) Time dependence of
$\chi_4(t)$ and $ \langle \frac{Q(t)}{N} \rangle $ at $T=0.60$. The long time
value of $\langle \frac{Q(t)}{N} \rangle = \rho V_a$ (solid line), 
where $V_{a} = \frac{4 \pi a^3}{3}$ corresponds to the
probability of finding a random overlapping particle.
The error bars in $Q(t)$ and $\chi_4(t)$ are calculated by averaging
over $100$ successive independent blocks, and are 
representative of the error in all points.}
\label{fig1} 
\end{figure}

We plot $g^{*}_4(r,t) $ for several $t$ at our second coldest
temperature $T=0.60$ in Figs.~\ref{fig1}(a) and (b). The positions of
the peaks in $g^{*}_4(r,t)$ are identical to the positions of the
peaks in $g(r)$ (not shown). We confirm that $g^{*}_4(r,t) = g(r) -1 $
in the ballistic and diffusive regime. Note that in the diffusive
regime $g^{\rm{ol}}_4(r,t)$ is the pair correlation function of the
random overlaps normalized by $\langle \frac{Q(t)}{N} \rangle^2$ to
yield $g(r)$.  We plot $\chi_4(t)$ and $\langle \frac{Q(t)}{N} \rangle
$ at $T=0.60$ in Fig.~\ref{fig1}(c); as found in
Refs.~\cite{dfpg,gns}, $\chi_4(t)$ has a maximum at an intermediate
time $t_4^{*}$.  Refs.~\cite{dfpg,gns} showed that this maximum
increases and shifts to longer time as $T$ decreases toward $T_{\rm
MCT}$, and we further find that $t_4^{*}$ is in the
$\alpha$-relaxation regime at each $T$. $g^{*}_4(r,t)$ deviates from
$g(r)-1$ when $\langle \frac{Q(t)}{N}\rangle $ deviates from unity and
$\chi_4(t)$ becomes non-zero\cite{grand}.  The amplitude and range of
$g^{*}_4(r,t)$ increase with increasing $t$ until a time $t_4^{*}$.
At $t_4^*$, $g^{*}_4(r,t^{*}_4)$ (indicated by the solid curve in
Figs.~\ref{fig1}(a,b)) exhibits a long tail which decreases slowly to
zero with increasing distance. For $t$ greater than $t_4^*$, the
amplitude and range of $g^{*}_4(r,t)$ decrease, and $g^{*}_4(r,t)= g(r) -1$
when $\chi_4(t)$ decays to zero (not shown). The
positions of the peaks in $g^{*}_4(r,t)$ do not appear to change with
decreasing $T$.

\begin{figure}[tbp]
\begin{center}
\includegraphics[clip,width=15cm]{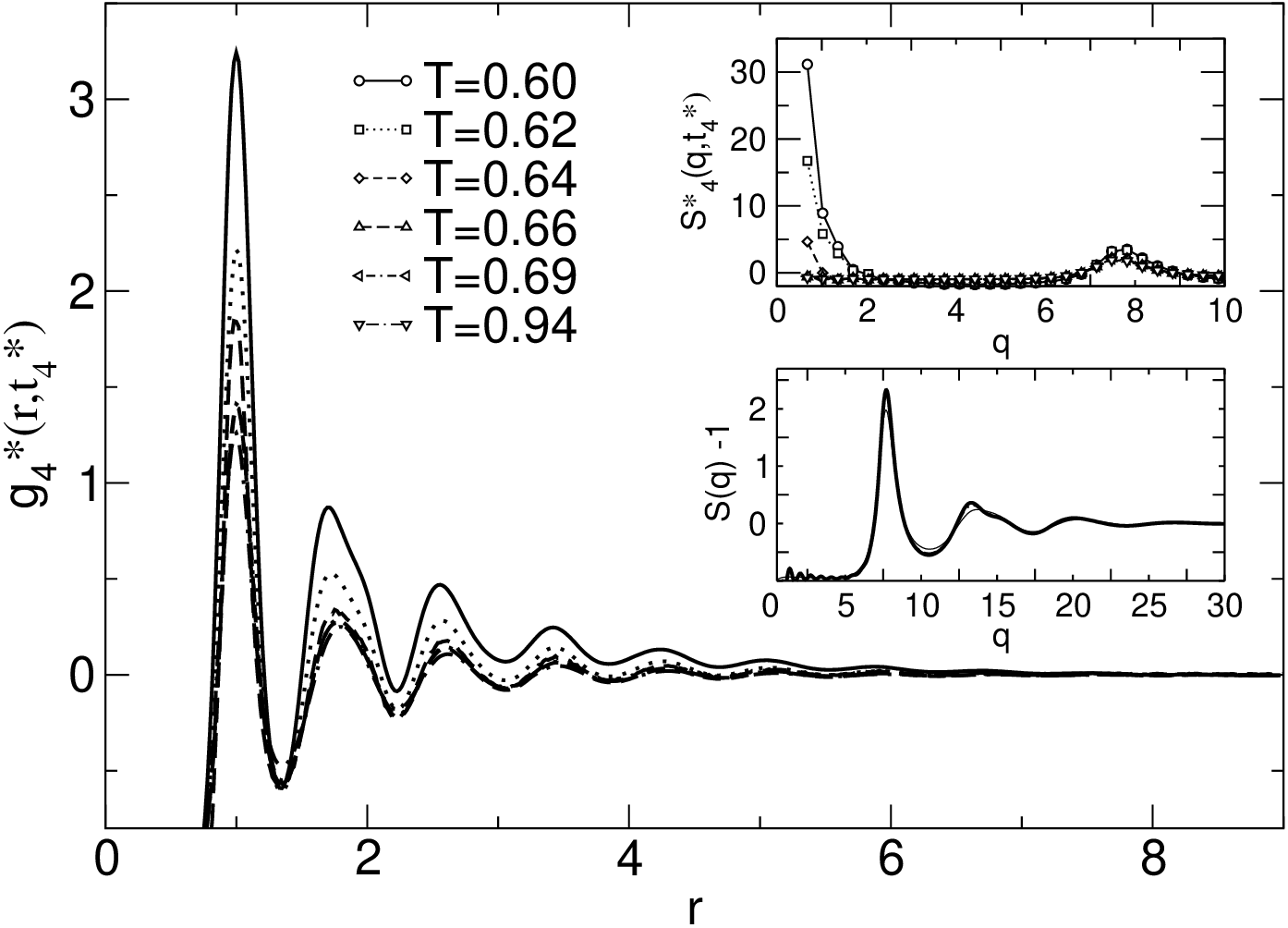}
\end{center}
\caption{ Temperature dependence of $g^{*}_4(r,t)$
at six values of $T$, indicated in legend. 
Insets show the structure factor $S^{*}_4(q,t^{*}_4)$ and static structure 
factor $S(q)$ for the same values of $T$. We note that $g^{*}_4(r,t)$ is
analogous to $g(r)-1$, and $S^{*}_4(q,t)$ is analogous to $S(q)-1$. }
\label{fig2}
\end{figure}

Fig.~\ref{fig2} shows the $T$ dependence of $g^{*}_4(r,t)$  at the peak 
characteristic time $t_4^{*}$ when the correlations at each $T$ are
most pronounced, as measured by $\chi_4(t)$. The inset of Fig.~\ref{fig2} shows the ``four-point'' 
structure factor $S^{*}_4(q,t_4^{*}) = \rho \int d\textbf{r} 
g^{*}_4(r,t^{*}_4) \sin(qr)/qr $, and static structure factor 
$S(q)-1 = \rho \int d\textbf{r} [g(r)-1] \sin(qr)/qr $.
We find that while $S(q)$ shows no change at small $q$, $S_4(q,t_4^{*})$
develops a peak at small q that grows with decreasing $T$, indicating a
growing range of correlations.

\begin{figure}[tbp]
\begin{center}
\includegraphics[clip,width=15cm]{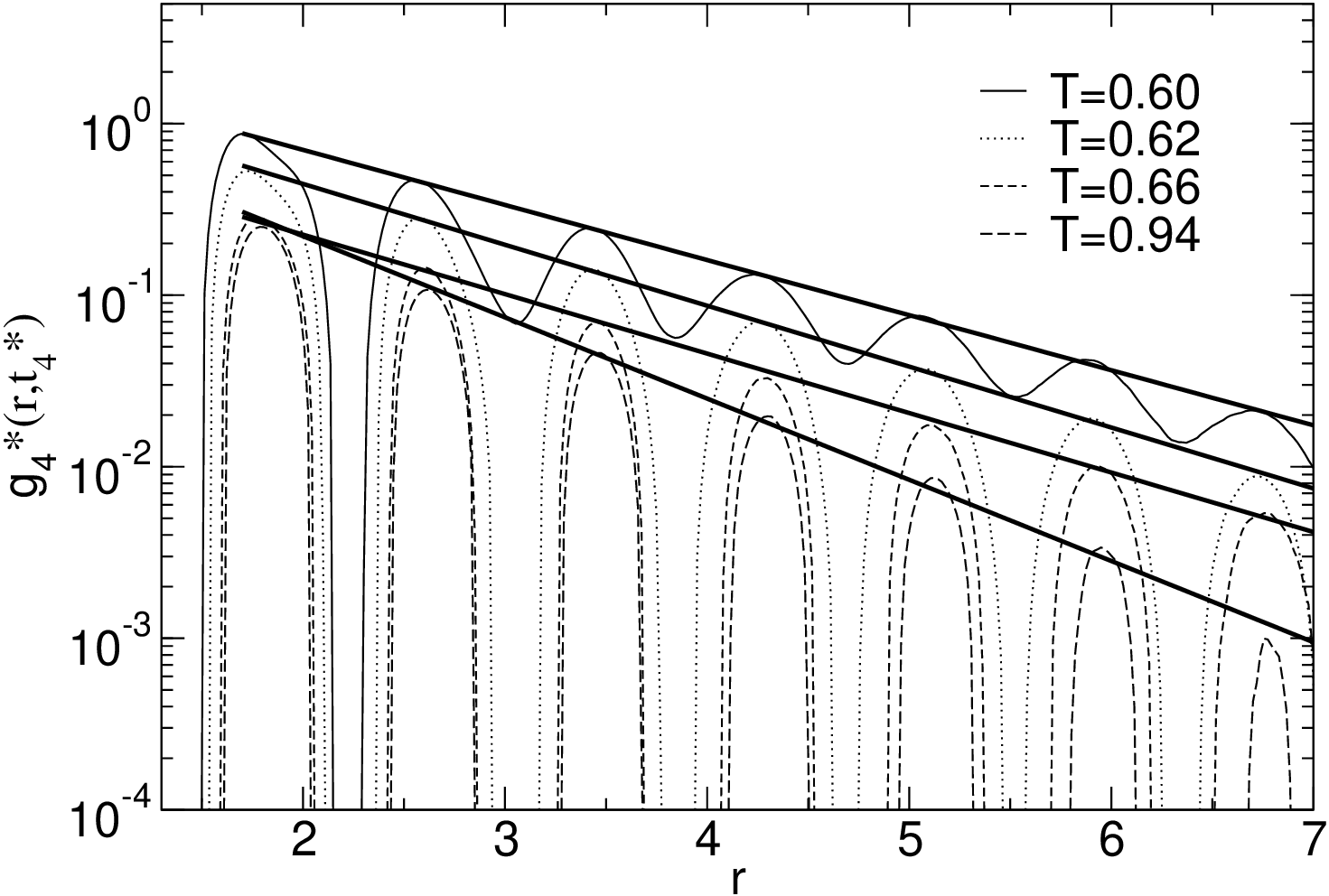}
\end{center}
\caption{ Semilogarithmic plot of $g^{*}_4(r,t^{*}_4)$ in the range
$1.7<r<7$ at four $T$. The solid lines are the exponential curves
obtained from an ``envelope fit''.}
\label{fig3}
\end{figure}

\begin{figure}[tbp]
\begin{center}
\includegraphics[clip,width=15cm]{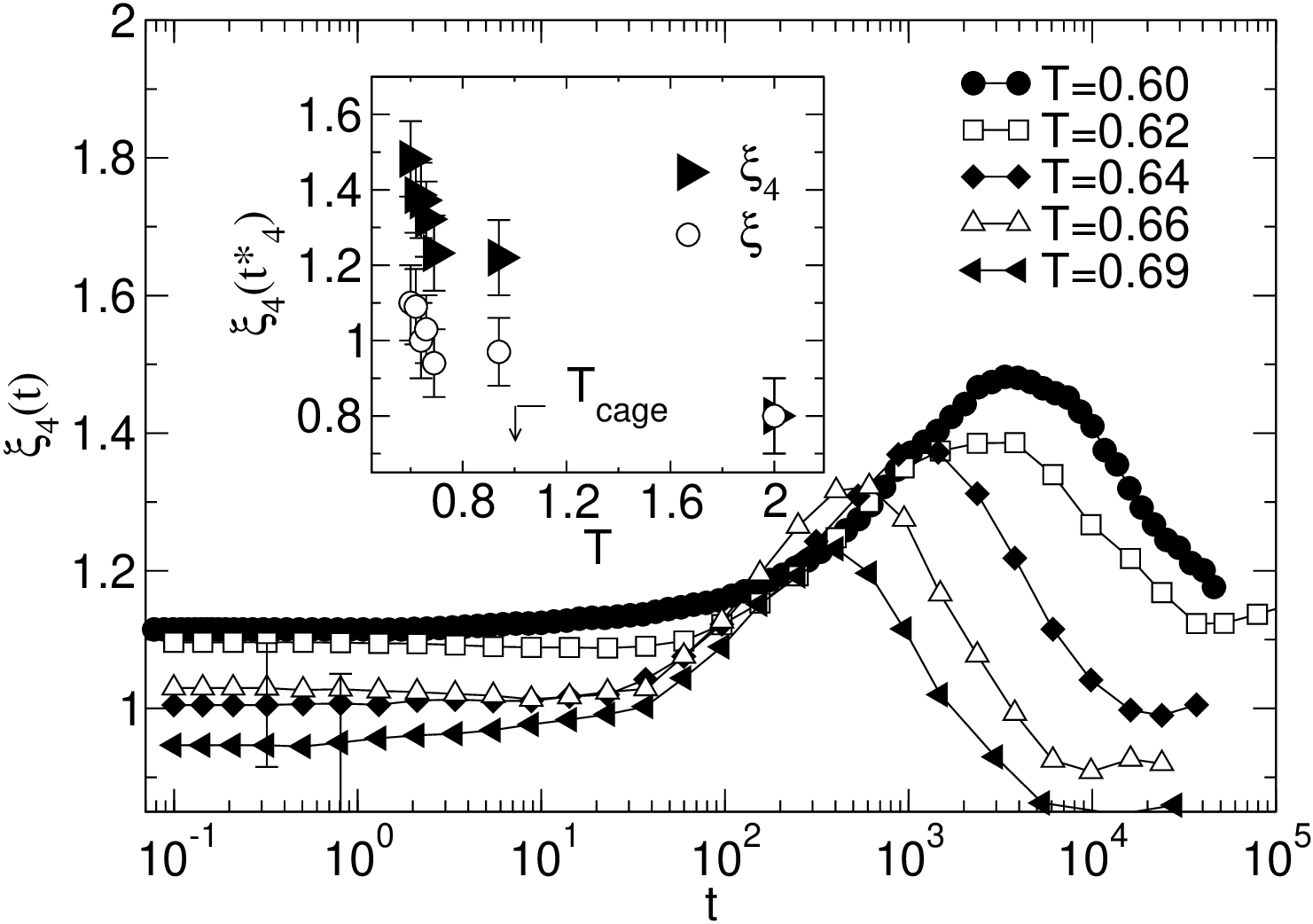}
\end{center}
\caption{ Time and $T$ dependence of $\xi_4(t)$. 
The inset shows $\xi_4(t_4^{*})$ vs. $T$. The data shown is smoothed
by performing running average over successive groups of five points.}
\label{fig4} 
\end{figure}

Fig.~\ref{fig3} provides a close-up of the behavior of
$g^{*}_4(r,t^{*}_4)$ for $1.7 < r < 7$, for several values of $T$. To
extract a value for $\xi_4(t)$, we fit peaks of $g^{*}_4(r,t)$ in the
range shown to the exponential function $y(r) =
a*\rm{exp}(-r/\xi_4(t))$. We refer to this method as an ``envelope
fit''.  The time dependence of $\xi_4(t)$ obtained from this fit is
plotted for several state points in Fig.~\ref{fig4}.  We see that the
qualitative behavior of $\xi_4(t)$ is similar to that of $\chi_4(t)$:
$\xi_4(t)$ has a maximum in time, and as $T$ decreases, the amplitude
and time of this maximum increase.

The length scale $\xi_4(t)$ characterizes the typical distance over
which ``overlapping'' particles are spatially correlated, and includes
contributions from the static two-point density correlations.  At
temperatures above the onset of caging, where the dynamics is
everywhere homogeneous, $\xi_4(t_4^*)$ and $\xi$ coincide. Below the
onset of caging, $\xi_4(t_4^*)$ begins to grow larger than $\xi$; over
the limited $T$-range of our simulations, we find that $\xi_4(t)$
increases from $0.8 \pm 0.1$ particle diameters above $T_{\rm cage}$
to $1.5 \pm 0.1 $ particle diameters within $5\%$ of $T_{{\rm MCT}}$
(Fig.~\ref{fig4} inset). In contrast, we find that $\xi$, which is
coincident with $\xi_4(t)$ at short times when $ \langle
\frac{Q(t)}{N} \rangle\ =1$, changes less, from $0.8 \pm 0.1$ to $1.1
\pm 0.1$ particle diameters. Thus, over the $T$-range studied, the
characteristic distance over which particle motion is most correlated
{\it grows to exceed the static correlation length}.  While we cannot
reliably predict the behavior of $\xi_4(t_4^*)$ at lower $T$, we find
no tendency for slowing down of its growth, unlike the low-$T$
behavior of $\xi$ which is known to remain finite and small.  We do
observe a slight depression of $\xi_4(t)$ and $\chi_4(t)$ at our
lowest $T=0.59$ (not shown), but we speculate this may be due to
finite size effects, which we will explore in detail in a later
publication.

The relatively small but growing correlation length calculated here
from the four-point spatiotemporal density correlation function should
be contrasted with that characterizing the size of highly mobile
regions within the fluid \cite{kdppg-gssg}.  That length was shown to
grow much more rapidly on cooling, approaching the size of the
simulation box close to $T_{\rm{MCT}}$.  Interestingly, whereas the
correlation length of highly mobile regions was found to be largest on
a time scale in the $\beta$-relaxation regime, the length calculated
in the present paper is largest in the $\alpha$-regime. We note that
$g_4(r,t)$ is dominated by ``caged'' particles, and thus $\xi_4(t)$
may be related to length scales calculated in
Ref.~\cite{mountain-das}. The relationship of the different length
scales characterizing dynamical heterogeneity will be explored in a
subsequent publication.

\begin{thebibliography}{1}
\bibitem{grand} Strictly speaking, a simulation in the grand
canonical ensemble would instead yield a small but non-zero $\chi_4(t)$
value at early times.
\end{thebibliography}
\end{document}